\documentclass[a4paper,14pt]{article}
\usepackage[utf8]{inputenc}
\usepackage[T2A]{fontenc}
\usepackage[english]{babel} 
\usepackage[linktocpage,unicode]{hyperref}
\usepackage{csquotes,graphicx,fancyhdr}
\usepackage{setspace,longtable,graphicx,hyphenat,hyperref,fancyhdr,ifthen,everypage,enumitem,amsmath}
\usepackage[dvipsnames]{xcolor}
\usepackage[export]{adjustbox}
\usepackage[linktocpage,unicode]{hyperref}
\hypersetup{unicode=true,colorlinks=true,linkcolor=black,citecolor=blue,urlcolor=blue}
\usepackage{url}
\usepackage{indentfirst}
\usepackage{amssymb,amsmath}
\usepackage{geometry}
\usepackage[style=gost-numeric,bibencoding=auto,backend=biber,babel=other,sorting=none,language=auto,autolang=other,doi=false,url=false]{biblatex}
\AtEveryBibitem{\clearfield{day}
	\clearfield{month}
	\clearfield{endday}
	\clearfield{endmonth}}
\bibliography{bibliography}
\usepackage{subcaption}
\usepackage{multido}
\newcommand{\e}{\text{e}}
\newcommand{\n}{\text{n}}
\newcommand{\D}{\text{D}}
\newcommand{\M}{\text{M}}
\newcommand{\N}{\text{N}}
\newcommand{\K}{\text{K}}

\newcommand{\CC}{\text{C}}
\newcommand{\LL}{\text{L}}
\newcommand{\Pl}{\text{Pl}}
\newcommand{\cc}{\text{c}}
\makeatletter
\ams@newcommand{\vardot}[2]{%
  {\mathop{#2\kern0pt}\limits^{\vbox to-1.4\ex@{\kern-\tw@\ex@
   \hbox{\normalfont\multido{}{#1}{.}}\vss}}}}
 \makeatletter
\def\@fnsymbol#1{\ensuremath{\ifcase#1\or \dagger\or  *\or \dagger\dagger
   \or \ddagger\ddagger \else\@ctrerr\fi}}
    \makeatother
\makeatother
\title{Self-tuning inflation}

\author{{Polina Petriakova}$^{1}$\thanks{e-mail: polinapetriakova@gmail.com} \and
{Sergey~G.~Rubin}$^{1,2}$\thanks{e-mail: sergeirubin@list.ru}}

\date{\em \small
$^1$ National Research Nuclear University MEPhI (Moscow Engineering Physics Institute), 115409, Russian Federation, Moscow, Kashirskoe shosse, 31\\
$^2$ N. I. Lobachevsky Institute of Mathematics and Mechanics, Kazan Federal University, 420008, Russian Federation, Kazan, Kremlevskaya street, 18}
\begin{document}
\maketitle
\begin{abstract}
We develop an inflationary model without small parameters on the basis of multidimensional $f(R)$ gravity with a minimally coupled scalar field. The model is described by two stages of space expansion. The first one begins at energy scales about the D-dimensional Planck mass and ends with the de Sitter metric of our space and the maximally symmetric extra dimensions. In the following, the quantum fluctuations produce a wide set of inhomogeneous extra metrics in causally disconnected regions quickly generated in the de Sitter space. We find a specific extra space metric that leads to the effective Starobinsky model that fits the observational data. 
\end{abstract}
\section{Introduction}
The rapid development of observational cosmology has imposed serious limitations on inflationary models \cite{2020A&A...641A..10P}. Despite this, a significant number of models do not contradict observations. Most models of inflation are based on the postulation of a specific form of scalar fields potential without clarification of its origin \cite{2014PDU.....5...75M}. Another promising approach to providing a viable inflationary stage without relying on scalar fields is modified gravity. The modifications of the Einstein–Hilbert action include higher-order curvature invariants that naturally appear due to the quantum corrections or in the string theory framework \cite{Birrell:1982ix}. 

It is often assumed that inflation takes place at high energies characterized by the Hubble parameter $H\sim 10^{13}$ GeV. This gives rise to the small parameter $H/M_{\Pl}\sim 10^{-6}$, which is characteristic of inflationary models. 
Another small parameter refers to the Lambda term, which must be at least much smaller than the Hubble parameter. In this paper, we discuss the way to avoid these defects. We elaborate an inflationary model based on $f(R)$ gravity acting in multidimensional space without the small parameters. 

The ability to avoid ghosts and the Ostrogradski instability distinguishes $f(R)$ theories from other higher-order gravity theories \cite{2007LNP...720..403W}. Inflationary scenarios have fueled 
additional interest in $f(R)$ theories \cite{2010PhRvD..81h4010B, 2017PhR...692....1N}, beginning with Starobinsky's seminal paper \cite{1980PhLB...91...99S}. This model has remained very successful up to now, and a large number of its generalizations with minor modifications have been built \cite{2010JCAP...06..005A,2011PhR...505...59N,2017JCAP...09..041M,2018PhRvD..98d3505L,2019PhRvD..99j4070O,2020PhLB..80535453C,2021PhRvD.104l4065O,2021PhRvD.103h3518G,2022JCAP...03..058I,2022PhRvD.105f3504R,2022NuPhB.97815779O}. Nevertheless, in the IR limit, the Starobinsky model implies that the cosmological constant must also be fine-tuned and there is no "natural" expectation for the unique coefficient in front of the $R^2$ term in the action to be large. 
The latter feature of the Starobinsky model can be resolved by considering it as a low-energy effective theory of a multidimensional theory due to the compactification of extra dimensions.  This approach was discussed in \cite{Asaka:2015vza}. Nevertheless, the fine tuning of the cosmological constant and the extra space stabilization problem remain. Internal subspaces should be static or nearly static, according to observations 
\cite{2008PhRvD..78h3527L,2015A&A...580A..22P}, because their dynamical behavior causes the fundamental physical constants to vary. In purely gravitational $f(R)$ models, only subspaces with negative curvature can be stabilized \cite{2002PhRvD..66d4014G,2003PhLB..576....5N}. The most popular way of reaching a stabilization with positive curvature is to invoke matter fields of non-geometric origin\footnote{However, there are other options, such as using dynamical spacetime theory \cite{2018PhRvD..98d3522B} or form-fields \cite{2003PhRvD..68d4010G}} 
or/and to extend the gravitational action by adding invariants, such as the Einstein-Gauss-Bonnet term \cite{Canfora:2016umq}, e.t.c. Nonetheless, positive curvature spaces in the EGB are only stable in a narrow range of parameters or are not stable at all, depending on the dimension of the subspace, whereas negative curvature subspaces are known to be stable in a wide range \cite{2021EPJC...81..136C}. 

The metric and the fields inside the horizon experience strong quantum fluctuations at high energies. These fluctuations can affect the classical dynamics and final states at low energies, including the extra space metric. For example, a $6${--}dimensional quadratic $f(R)$ model with the scalar field and the inhomogeneous static compact extra dimensions was considered in \cite{2020EPJC...80..970B}.

The possible influence of the matter field on the metric of extra dimensions is investigated in the framework of D-dimensional $f(R)$ gravity with a minimally coupled scalar field. We distinguish two stages of expansion: high and low energy. At the high energy stage, which is unobservable, the extra subspace is absolutely symmetrical and the scalar field slowly rolls to the minimum, which is reached at $H \neq 0$. The extra space size evolves along with the scalar field and grows up to a certain size. At the same time, the number of causally disconnected areas is growing rapidly. They continue to multiply efficiently even when the minimum of potential of the scalar field is reached, since in this case, $H \neq 0$. This ends the first stage of the evolution of the Universe, which at the moment is one of the causally disconnected regions. In some such regions, the symmetry of the extra space metric is violated due to the fluctuation of the scalar field. As a result, the specific deformed configuration of the internal subspace provides the second stage, which is Starobinsky’s effective model. We have obtained the effective values of the model without involving unnatural parameters, i.e., larger or smaller than the order of magnitude.

This paper is organized as follows: Section \ref{Outlook} contains the basic equations of our model with the chosen general metric. In Section \ref{HES} we consider the dynamics of the D{-}dimensional metric with the scalar field at extremely high energies. We show that the extra absolutely symmetrical space size and the Hubble parameter are constants at the end of the first stage. Section \ref{LES} is devoted to a description of the inhomogeneous extra space. We show analytically that in the case of a homogeneous distribution of the scalar field, there exists only the maximally symmetrical metric. We present several numerical solutions with scalar field fluctuations that disturb the maximally symmetrical extra space metric. Among the found solutions, we choose an appropriate metric suitable for describing the effective Starobinsky model and the observed data. Conclusions are made in Section \ref{Conclusion}.

\section{Outlook}\label{Outlook}
Consider the $f(R)$ theory with a minimally coupled scalar field $\varphi$ in $\D = 4 + \n$ dimensions 
\begin{equation}\label{S0}
S = \frac{m_{\D}^{\D-2}}{2}  \int d^{\D} x \sqrt{|g_{\D}|} \,  \Bigl( f(R) + \partial^{\M}\varphi \, \partial_{\M}\varphi -2 V\bigl( \varphi \bigr) \Bigr)\, ,
\end{equation}
where $g_{\D} \equiv \text{det} g_{\M\N}$, $\M,\N =\overline{1,\D}$ and $f(R)$ is a smooth function of the D-dimensional Ricci scalar $R$. The metric of $M_1 \times M_3 \times M_\n$ manifold is chosen in the form 
\begin{equation}\label{metric_common}
    ds^{2} = dt^2 - \e^{2 \alpha (t)}\delta_{ij}dx^i dx^j - \e^{2 \beta (t)}m^{-2}_{\D} \biggl(\bigl(d x^4\bigr)^2 +r^2(x^4) \bigl(d x^5\bigr)^2 + ... + r^2(x^4) \prod\limits_{\text{k}=5}^{\D-2} \Bigl(\sin^2{x^{\text{k}}}\Bigr) \bigl(d x^{\text{k}+1}\bigr)^2 \biggr).
\end{equation}
Variation of action \eqref{S0} with respect to the metric $g^{\M\N}$ and scalar field $\varphi$ leads to the known equations
\begin{align}\label{eqMgravity}
-\frac{1}{2}{f}(R)\delta_{\N}^{\M} + \Bigl(R_{\N}^{\M} +\nabla^{\M}\nabla_{\N} &- \delta_{\N}^{\M} \Box_{\D} \Bigr) {f}_R = - T_{\N}^{\M},  \\ 
\label{eqMscalarfield} \Box_{\D} \, \varphi + V^{\prime}_{\varphi} &=0 \end{align}
with $f_R = \dfrac{df(R)}{dR}$, $\Box_{\D}= \nabla^{\M} \nabla_{\M}$ and $V^{\prime}_{\varphi} = \dfrac{d V\bigl( \varphi \bigr)}{d\varphi}$. The arbitrary potential satisfies conditions $\left.V \bigl(\varphi \bigr) \right|_{\varphi =0}=0$, $\left. V'_{\varphi} \bigl(\varphi \bigr)\right|_{\varphi =0}=0$. The corresponding stress-energy tensor of the scalar field $\varphi$ is
\begin{equation}
    T_{\N}^{\M} = \frac{\partial L_{matter}}{\partial\bigl(\partial_{\M} \varphi \bigr)}\partial_{\N}\varphi - \frac{\delta_{\N}^{\M}}{2} L_{matter} = \partial^{\M}\varphi \, \partial_{\N}\varphi - \frac{\delta_{\N}^{\M}}{2} \, \partial^{\K}\varphi \, \partial_{\K}\varphi +  \delta_{\N}^{\M}V\bigl( \varphi \bigr) \, .
\end{equation}
We use the following conventions for the curvature tensor $R_{\M\N\K}^\LL=\partial_\K\Gamma_{\M\N}^\LL-\partial_\N \Gamma_{\M\K}^\LL +\Gamma_{\CC\K}^\LL\Gamma_{\N\M}^\CC-\Gamma_{\CC\N}^\LL \Gamma_{\M\K}^\CC$
and the Ricci tensor $R_{\M\N}=R^\K_{\M\K\N}$. Equation \eqref{eqMscalarfield} is known to be the consequence of equations \eqref{eqMgravity}.

\section{The first stage. High-energy space expansion}\label{HES}

In this section, we study the dynamics of the field and the $\D-$dimensional metric at extremely high energies, starting from the scale of the order of the $m_\D<m_{\Pl}$. We show that the extra space size and the Hubble parameter do not depend on time at the end of this stage. 

Let us consider metric \eqref{metric_common}, where $M_\n$ is the n-dimensional sphere, i.e. $r(x^4)=\sin{x^4}$, and all dynamical variables depend only on time.
In this case, the Ricci scalar is
\begin{equation}\label{Ricci_n_dim}
R(t)= 6\ddot{\alpha} + 2\n\, \ddot{\beta} +12 {\dot{\alpha}}^2 + \n \bigl( \n +1 \bigr) {\dot{\beta}}^2 + 6 \n \, \dot{\alpha} \dot{\beta} + \n \bigl(\n-1 \bigr)\,m^{2}_{\D} \e^{-2 \beta(t)} .
\end{equation}
Then, for the case where the scalar field depends only on time, $\varphi=\varphi(t)$, system \eqref{eqMgravity} for $(tt)$, $(x^1 x^1)=(x^2 x^2)=(x^3 x^3)$, $(x^4 x^4)=...=(x^{\D-1} x^{\D-1})${--}components and  \eqref{eqMscalarfield} take the following form:  
\begin{eqnarray}\label{eq_tt}
&& - \left(3\dot{\alpha} +\n\dot{\beta}\right)\dot{R} f_{RR} +\left( 3\ddot{\alpha} + \n\ddot{\beta} +3{\dot{\alpha}}^2  +\n{\dot{\beta}}^2 \right) f_{R} - \dfrac{f(R)}{2} = - \dfrac{\dot{\varphi}^2}{2} - V\bigl( \varphi \bigr)
, \\ \label{eq_xx}
&& \quad {\dot{R}}^2 f_{RRR} +\left(\ddot{R} + \bigl( 2 \dot{\alpha} + \n \, \dot{\beta}\bigr)\dot{R} \right) f_{RR} \, - \left( \ddot{\alpha} +3{\dot{\alpha}}^2 + \n \dot{\alpha} \dot{\beta} \right) f_{R} + \, \dfrac{f(R)}{2} = - \dfrac{\dot{\varphi}^2 }{2} + V\bigl( \varphi \bigr)\, ,  \\
\label{eq_thetatheta}
&& \quad{\dot{R}}^2 f_{RRR} +\left( \ddot{R} + \bigl(3\dot{\alpha} + \bigl(\n-1 \bigr)\dot{\beta}\bigr)\dot{R} \right) f_{RR} -\left( \ddot{\beta} +3\dot{\alpha} \dot{\beta} + \n {\dot{\beta}}^2 + \bigl(\n-1 \bigr)m^{2}_{\D}\e^{-2 \beta(t)} \right)f_{R} + \\ \nonumber
&& \qquad +  \, \dfrac{f(R)}{2} =  - \dfrac{\dot{\varphi}^2 }{2} + V\bigl( \varphi \bigr)\, , \\ 
\label{eq_scalar}
&& \quad \ddot{\varphi} + \Bigl(3\dot{\alpha}+\n\dot{\beta}\Bigr)\dot{\varphi} + V^{\prime}_{\varphi} =0,
\end{eqnarray} 
where the dot notation refers to the time derivatives. It is of interest to study the asymptotic behavior of the metric and the scalar field with the help of equations \eqref{eq_tt}{--}\eqref{eq_scalar}. They have the form\footnote{In order for $m^{2}_{\D}\e^{-2\beta_{\cc}}>0$, for any sign of $a$, the inequality $c<0 $ must hold.}: $\varphi_{as}=0$ and
\begin{align}
\label{Hubble}
H_{as}^2 &= \dfrac{- \bigl(\n+2\bigr)\pm \sqrt{\bigl(\n+2\bigr)^2-4a\n \bigl(\n+4\bigr)c\, }}{6a\n \bigl(\n+4\bigr)}\, ; \\ \label{beta}
\e^{-2\beta_{as}} &=  \dfrac{- \bigl(\n+2\bigr)\pm \sqrt{\bigl(\n+2\bigr)^2-4a\n \bigl(\n+4\bigr)c \, }}{2a\n \bigl(\n+4\bigr)\bigl(\n-1\bigr)\,m^{2}_{\D}} \equiv \e^{-2\beta_{\cc}} \end{align}
for a specific form of the $f (R)$ function
\begin{equation}\label{fR}
  f (R) = aR^2 + R+ c .
\end{equation}
In the slow-roll approximation, $\dot{\varphi}^2 \ll V\bigl(\varphi \bigr)$, $|\ddot{\varphi}|\ll|V^{\prime}_{\varphi}|$, $\dot{\alpha} \simeq \text{const} \equiv H$ and $\ddot{\beta}\simeq \dot{\beta}\simeq 0$, the system of equations above is strongly simplified in the same way as for the standard inflationary scenario:
\begin{align}\label{eq_tt_appr}
 3H^2 f_{R} - \dfrac{f(R)}{2} & \simeq - V\bigl( \varphi \bigr), \\ 
\label{eq_thetatheta_appr}
- \bigl(\n-1 \bigr)\,m^{2}_{\D}\e^{-2 \beta_{\cc}} f_{R} +  \dfrac{f(R)}{2} & \simeq V\bigl( \varphi \bigr) ,\\ 
\label{eq_scalar_appr}
3H\dot{\varphi} + V^{\prime}_{\varphi} & \simeq  0 .
\end{align}
Here the Ricci scalar is $R \simeq 12 H^2  + \n \bigl(\n-1 \bigr)\,m^{2}_{\D}\e^{-2 \beta_{\cc}}$.
In the absence of extra dimensions and for the standard linear gravity, $f(R) = R$, equations \eqref{eq_tt_appr} and \eqref{eq_scalar_appr} yield the well-known relations
\begin{equation}\label{stand_infl_4dim}
    3H^2 \simeq \left. V \bigl(\varphi \bigr) \right|_{m_{\text{Pl}}=1} \qquad \text{and} \qquad 3H\dot{\varphi} \simeq - V^{\prime}_{\varphi}.
\end{equation}
Equations \eqref{eq_tt_appr} and \eqref{eq_thetatheta_appr} lead to the following relationship:
\begin{equation}\label{beta_Hubble_ratio}
m^{2}_{\D}\e^{-2 \beta_{\cc}} = \dfrac{3H^2}{\bigl(\n-1 \bigr)},
\end{equation}
valid for any form of $f(R)$ and $V(\varphi)$.  As can be seen, in the limit $H \rightarrow 0$, corresponding to present times, expression \eqref{beta_Hubble_ratio} leads to an infinitely large size of extra dimensions $\e^{\beta_{\cc}} $, which obviously contradicts observations. The static inhomogeneous extra dimensions can resolve this limitation, as will be shown in the next section.
\begin{figure}[!ht]
\centering
\begin{subfigure}[t]{0.3\linewidth}
\includegraphics[width=\linewidth]{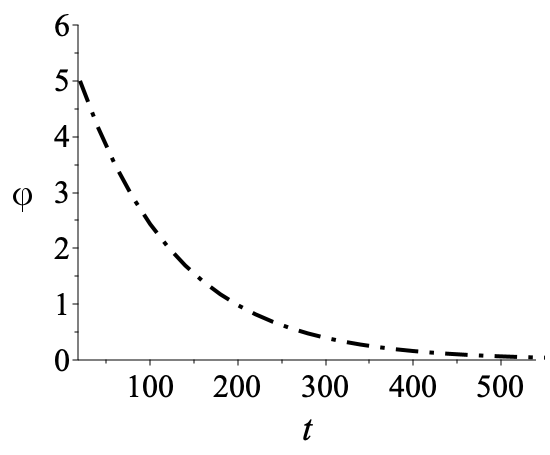}
\end{subfigure}
\hspace{0.35cm}
\begin{subfigure}[t]{0.3\linewidth}
\includegraphics[width=\linewidth]{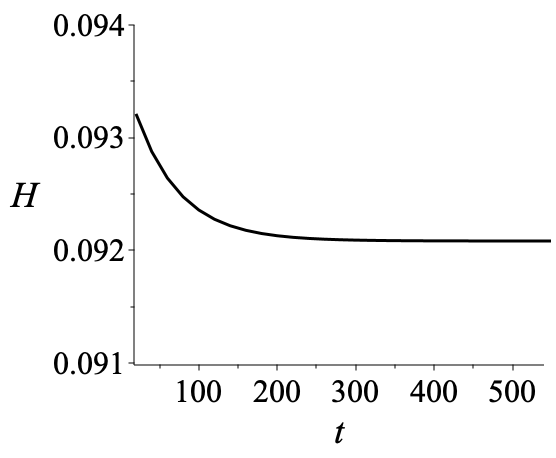}
\end{subfigure}
\hspace{0.35cm}
\begin{subfigure}[t]{0.3\linewidth}
\includegraphics[width=\linewidth]{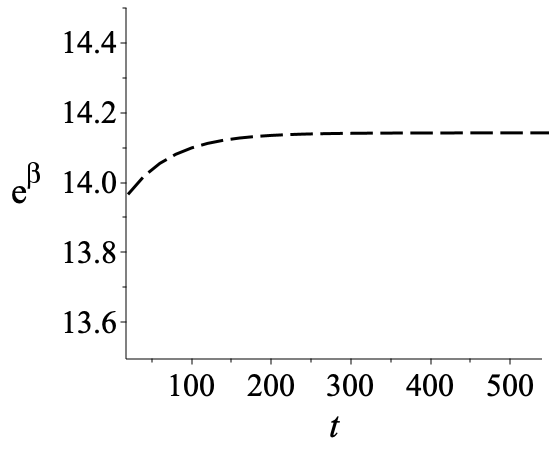}
\end{subfigure}
\caption{The first-stage behavior of the scalar field $\varphi(t)$ and the Hubble parameter $H(t)$, as well as the size of the extra space $\e^{\beta(t)}$ for the parameters used below $a=20m_{\D}^{-2}$, $c=-0.95m_{\D}^{2}$, $\n=6$, $m=0.05m_{\D}$.}
\label{beta_Hubble}
\end{figure}

Since the solution of equation \eqref{eq_scalar_appr} reveals a decreasing scalar field with time, the Hubble parameter will slowly decrease with time, tending to a constant value. The size of the extra space increases with time and also tends to a constant value. The behavior of the scalar field $\varphi(t)$, the Hubble parameter $H(t)$, and the size of the extra subspace $\e^{\beta(t)}$ are shown in Figure \ref{beta_Hubble} in units of $m_\D=1$.

The first stage of the space expansion is finished at $ V\bigl(\varphi_{as}\bigr)=0$ with conditions \eqref{Hubble}, \eqref{beta}. After integration over the extra coordinates\footnote{$\int d^{\n} x \sqrt{|{g}_{\n}|}=\dfrac{2\pi^{\tfrac{\n+1}{2}}m^{-\n}_{\D}\e^{\n\beta_{\cc}}}{\,\Gamma\left(\tfrac{\n+1}{2}\right)} \equiv \upsilon_{\n}\,m^{-\n}_{\D}\e^{\n\beta_{\cc}}$} using $f(R) \simeq \frac{1}{2} f_{RR}(R_\n) R_4^2 + f_{R}(R_\n) R_4 + f(R_\n)$ decomposition, effective action \eqref{S0} takes the form
\begin{equation}\label{S_eff}
S^{\,I}_{eff} = \dfrac{m_{\Pl}^{2}}{2} \int d^4 x \sqrt{|g_4|} \Bigl(a_{eff}R_4^2 + R_4 + c_{eff} \Bigr).
\end{equation}
Here and in the following, we assume the quadratic form of $f(R)$ as in \eqref{fR}.
The effective parameters and the relationship between the 4-dimensional and the $\D$-dimensional Planck masses
\begin{equation}\label{coeff_eff_symm}
a_{eff} m^2_{\Pl} = \dfrac{1}{2} \, \upsilon_{\n} m_{\D}^{2}\e^{\n\beta_{\cc}} f_{RR}\bigl(R_{\n}\bigr), \qquad  m^2_{\Pl} = \upsilon_{\n} m_{\D}^{2}\e^{\n\beta_{\cc}} f_R\bigl(R_{\n}\bigr), \qquad c_{eff}m^2_{\Pl} =  \upsilon_{\n} m_{\D}^{2}\e^{\n\beta_{\cc}} f\bigl(R_{\n}\bigr)
\end{equation}
are related to the Ricci scalar $R_{\n}=\n \bigl(\n-1 \bigr)\,m^{2}_{\D} \e^{-2 \beta_{\cc}}$ of the extra dimensions. The Hubble parameter equals a constant according to \eqref{Hubble}. It is the non-zero value of $c_{eff}$ that causes the de Sitter metric of our 4-dimensional space. 

Now let us perform some numerical estimations. The parameter values $a=20 m^{-2}_\D$, $c=-0.95m^{2}_\D$, and the number of extra dimensions $\n=6$ are suitable for our purposes. In this case, relations \eqref{coeff_eff_symm} give the effective parameter values $a_{eff} m^2_{\Pl} \simeq  5.29 \cdot 10^9$, $c_{eff}m^2_{\Pl} \simeq -9.26  \cdot 10^7 m^{4}_{\D}$. The extra space radius and the Hubble parameter are $\e^{\beta_{\cc}} \simeq 14.14 m^{-1}_{\D}$ and $H\simeq 0.092 \,  m_{\D}$ correspondingly. The D-dimensional Planck mass relates to the 4-dimensional one as 
$ m^2_{\Pl} \simeq 1.85 \cdot 10^{9} m^2_\D$ according to \eqref{coeff_eff_symm}, which gives $m_\D \sim 10^{14}$~GeV. Now, the Hubble parameter and the extra space size can be expressed in physical units\footnote{$m_{\Pl}=M_{\Pl}/\sqrt{8\pi} = 2.4 \cdot 10^{18}$ GeV.}: $H \sim 10^{13}$~GeV and $\e^{\beta_{\cc}} \sim 10^{-27}$~cm.

The first, high energy stage of the space expansion is finished. The space expansion and the metric or scalar field fluctuations can last for an arbitrarily long time. Their scale quickly overcomes the present horizon. The fluctuations within the extra space are of most interest. As was shown in \cite{2021arXiv210908373R}, some of them can deform the extra dimensions significantly. It leads to an alternation of the Lagrangian parameters $a_{eff}$ and $c_{eff}$ and launches the observable inflation.

\section{The second stage. Observable inflation}\label{LES}

\subsection{Inhomogeneous compact extra dimensions}
The high-energy state is finished with the de-Sitter 4-dimensional metric and maximally symmetrical extra space, see formulas \eqref{Hubble}-\eqref{beta}. The fluctuations deform extra metric in the causally disconnected regions (the pocket universes in the future). In this section, we discuss the way to choose a pocket universe with an appropriate metric of extra space. The equations of motion appear to be too complicated to be solved without any approximation. Fortunately, the inflationary stage usually assumes the slow variation of dynamical variables. In particular, the Hubble parameter $H\simeq\text{const}\sim 10^{13}$ GeV and we widely use this fact in the following to find the metric of the extra space depending on $H$.

The scalar field fluctuations at the de Sitter stage can break the maximally symmetrical extra space metric \cite{2021arXiv210908373R}, which is the reason for an inhomogeneous metric formation \cite{2015EPJC...75..333R}. Here we consider an inhomogeneous n-dimensional extra metric
\begin{equation}\label{metric_deformed_r}
    ds^{2} = dt^2 - \e^{2Ht}\Bigl(\bigl(d x^1\bigr)^2 + \bigl(d x^2\bigr)^2 + \bigl(d x^3\bigr)^2\Bigr) - \e^{2 \beta_{\cc}}m^{-2}_{\D} \Bigl(du^2 + r^2(u)\,d\Omega_{\n-1}^2\Bigr)\,
\end{equation} 
with the renaming of the coordinate $x^4 \equiv u$ in \eqref{metric_common}. The factor $\e^{2\beta_\cc}$ is taken from \eqref{beta} for convenience. The Hubble parameter $H$ and the metric function $r(u)$ are defined below. The Ricci scalar
\begin{equation}\label{Ricci_n_dim_deformed_r}
R(u)= 12 H^2 - \bigl(\n-1 \bigr) \left( \dfrac{2 r^{\prime\prime}}{r}\,+ \bigl(\n-2 \bigr) \left(\dfrac{r^\prime}{r}\right)^2\,- \dfrac{(\n-2)}{r^2} \right)m^{2}_{\D} \e^{-2 \beta_{\cc}}
\end{equation}
does not depend on time. Remind that our aim is to find the effective 4-dimensional inflationary model and fix appropriate values of the Lagrangian parameters. For this purpose, we use the fact of slow motion of the dynamical variables and put approximately $H=\text{const}, \varphi=\varphi(u), \beta=\beta_\cc$ throughout this section. The notation used are $\prime \equiv d / du$ and $\prime\prime \equiv d^2 / du^2$ respectively.
Then equations \eqref{eqMgravity} for $(tt)=...=(x^3 x^3)$, $(uu)$ and $(x^5 x^5)=...=(x^{\D-1} x^{\D-1})${--}components and \eqref{eqMscalarfield} become
\begin{align}\label{eq_tt_deformed_r}
& \biggl(\bigl(R^{\prime}\bigr)^2 f_{RRR} +\Bigl(R^{\prime \prime} + \bigl(\n-1 \bigr) \dfrac{r^\prime}{r}R^{\prime} \Bigr)f_{RR} \biggr)m^{2}_{\D}\e^{-2\beta_{\cc}} +  
3H^2 f_{R} - \dfrac{f(R)}{2} = - \dfrac{\bigl(\varphi^{\prime}\bigr)^2}{2} m^{2}_{\D}\e^{-2\beta_{\cc}} - V\bigl( \varphi \bigr), \\
\label{eq_thetatheta_deformed_r}
& \bigl(\n-1 \bigr)\biggl(\dfrac{r^\prime}{r}\,R^{\prime}\,f_{RR} - \dfrac{r^{\prime\prime}}{r} \, f_R \biggr)m^{2}_{\D}\e^{-2\beta_{\cc}} - \, \dfrac{f(R)}{2} =  \dfrac{\bigl(\varphi^{\prime}\bigr)^2}{2} m^{2}_{\D}\e^{-2\beta_{\cc}} - V\bigl( \varphi \bigr) \,, \\
\label{eq_phiphi_deformed_r}
& \Biggl(\bigl(R^{\prime}\bigr)^2 f_{RRR} \, + \biggr(R^{\prime \prime}+\bigl(\n-2 \bigr)\dfrac{r^\prime}{r}\,R^{\prime} \biggl)f_{RR}\Biggr)m^{2}_{\D}\e^{-2\beta_{\cc}} - \\ \nonumber
& \qquad - \biggl(\dfrac{r^{\prime\prime}}{r} + \bigl(\n-2 \bigr)\left(\dfrac{r^\prime}{r}\right)^2 - \dfrac{(\n-2)}{r^2} \biggr) m^{2}_{\D}\e^{-2\beta_{\cc}} f_R  - \, \dfrac{f(R)}{2} = - \dfrac{\bigl(\varphi^{\prime}\bigr)^2}{2}m^{2}_{\D} \e^{-2\beta_{\cc}} - V\bigl( \varphi \bigr) \, , \\
\label{eq_scalar_deformed_r}
& \biggl(\varphi^{\prime \prime} + \bigl(\n-1\bigr) \dfrac{r^\prime}{r}\,\varphi^{\prime} \biggr)m^{2}_{\D}\e^{-2\beta_{\cc}} - V^{\prime}_{\varphi} =0.
\end{align} 
There are three independent equations for three unknowns $H$, $r(u)$ and $\varphi(u)$. Time dependence of the Hubble parameter is neglected because of the slow rolling during the inflation.
Also, we will use the definition of the Ricci scalar
\eqref{Ricci_n_dim_deformed_r}
as the additional unknown function to avoid  3rd and 4th order derivatives in the equations\footnote{Note that equations \eqref{Ricci_n_dim_deformed_r}, \eqref{eq_thetatheta_deformed_r}, \eqref{eq_phiphi_deformed_r}, \eqref{eq_scalar_deformed_r} are not independent: $\biggl(\dfrac{d}{du}$\eqref{eq_thetatheta_deformed_r}$+\dfrac{f_R}{2} \cdot \dfrac{d}{du}$\eqref{Ricci_n_dim_deformed_r}$ +\varphi^\prime \cdot$\eqref{eq_scalar_deformed_r}$\biggr) \cdot \dfrac{r}{r^\prime (\n-1)} \, +$\eqref{eq_thetatheta_deformed_r}$=$\eqref{eq_phiphi_deformed_r}.}.
The constraint equation is obtained by
combining $\Bigl($\eqref{eq_tt_deformed_r}$-$\eqref{eq_phiphi_deformed_r}$\Bigr)\cdot\bigl(\n-1\bigr)-$\eqref{eq_thetatheta_deformed_r}$-$\eqref{Ricci_n_dim_deformed_r}$\cdot f_R$
\begin{equation}\label{constraint_simple}
    \Bigl( 3\bigl(\n+3 \bigr)H^2 - R \Bigr)f_R + \frac{f(R)}{2} =  - \dfrac{\bigl(\varphi^{\prime}\bigr)^2}{2} m^{2}_{\D}\e^{-2\beta_{\cc}} + V\bigl( \varphi \bigr) .
\end{equation}
It has only the first-order derivatives and can be used as a restriction on the boundary conditions of the coupled second-order differential equations.

Let us check that equations \eqref{eq_tt_deformed_r}{--}\eqref{eq_phiphi_deformed_r} lead to a maximally symmetric metric if the scalar field distribution is homogeneous. In this case, the combination of equations $\Bigl($\eqref{eq_tt_deformed_r}$-$\eqref{eq_phiphi_deformed_r}$\Bigr)\cdot\bigl(\n-1 \bigr)-$\eqref{eq_thetatheta_deformed_r}$-$\eqref{Ricci_n_dim_deformed_r}$\cdot f_R$ leads to the algebraic equation
\begin{equation}\label{integrable}
    \Bigl( 3\bigl(\n+3 \bigr)H^2 - R \Bigr)f_R + \frac{f(R)}{2} = V\bigl( \varphi \bigr).
\end{equation}
Therefore, there is a solution of constant curvature $R=\text{const}\equiv R_\cc$ for any form of $f(R)$ function in the absence of matter/constant scalar field. For~constant curvature the difference of equations \eqref{eq_tt_deformed_r}$-$\eqref{eq_thetatheta_deformed_r} allows to find 
analytically the function $r(u)$:
\begin{equation}\label{sin_r_u}
    r(u) = m_{\D} \e^{-\beta_{\cc}}\dfrac{\sqrt{(\n-1)}}{\sqrt{3}H}\sin\biggl(\dfrac{\sqrt{3}H}{\sqrt{(\n-1)}} \e^{\beta_{\cc}}m^{-1}_{\D} u\biggr)
\end{equation}
$\forall f_{R}(R_\cc) \neq 0$ and then $R_\cc = 12H^2 + 3\n H^2$ from \eqref{Ricci_n_dim_deformed_r}. Metric \eqref{metric_deformed_r} with \eqref{sin_r_u} of the extra space is the metric of an n-dimensional sphere. Our reasoning does not dependent on the specific form of the $f(R)$ function. Solution \eqref{sin_r_u} reduces system \eqref{eq_tt_deformed_r}-\eqref{eq_phiphi_deformed_r} to the previously examined case \eqref{eq_tt_appr}-\eqref{eq_thetatheta_appr}. Note that the solution to \eqref{integrable} is $f(R)= C_0 \sqrt{R -  3\bigl(\n+3 \bigr)H^2 \, }$  if $V\bigl( \varphi \bigr) =0$. This solution with $C_0=2 \sqrt{3H^2 \,}\Bigl( 6 a \bigl(\n+4 \bigr)H^2 +1 \Bigr) $ is equivalent to relation \eqref{Hubble} for chosen function \eqref{fR} and constant curvature $R = R_\cc$. 

The next subsection is devoted to the numerical solution of equations \eqref{Ricci_n_dim_deformed_r}{--}\eqref{eq_scalar_deformed_r}. Here we suppose that the functions $r(u),\varphi(u), R(u)$ and the variable $H$ are known. After integration over extra coordinates\footnote{$\int d^{\n} x \sqrt{|g_{\n}|}= m^{-1}_{\D}\e^{\beta_{\cc}}\, \int  du \int d^{\n-1} x \sqrt{|g_{\n-1}|} =  \dfrac{2\pi^{\tfrac{\n}{2}}m^{-\n}_{\D}\e^{\n\beta_{\cc}}}{\,\Gamma\left(\tfrac{\n}{2}\right)} \int\limits_{u_{\text{min}}}^{u_{\text{max}}} r^{\n-1} (u)\, du \equiv \upsilon_{\n-1}\,m^{-\n}_{\D}\e^{\n\beta_{\cc}} \int\limits_{u_{\text{min}}}^{u_{\text{max}}}  r^{\n-1} (u)\, du$.} using decomposition $f(R) = \frac{1}{2} f_{RR}(R_\n) R_4^2 + f_{R}(R_\n) R_4 + f(R_\n)$, action \eqref{S0} turns to the effective theory 
\begin{equation}\label{S_eff2}
S^{\, II}_{eff} = \dfrac{m_{\Pl}^{2}}{2} \int d^4 x \sqrt{|g_4|} \Bigl(a_{eff}R_4^2 + R_4 + c_{eff} \Bigr)
\end{equation}
for a specific form of the function $f(R)$ as \eqref{fR}. Here, the effective values of the parameters are determined by the following expressions:
\begin{align}\label{a_eff_deformed}
a_{eff} & = \upsilon_{\n-1} \e^{\n\beta_{\cc}} \frac{m_{\D}^{2}}{2m_{\Pl}^{2}} \int\limits_{u_{\text{min}}}^{u_{\text{max}}} f_{RR}\bigl(R_{\n}(u)\bigr) \, r^{\n-1} (u)\, du ,\\
\label{m_pl_eff_deformed}
m^2_{\Pl} & = \upsilon_{\n-1}m_{\D}^{2} \e^{\n\beta_{\cc}} \int\limits_{u_{\text{min}}}^{u_{\text{max}}}f_R\bigl(R_{\n}(u)\bigr) \, r^{\n-1} (u)\, du ,\\ \label{Lambda_eff_deformed}
c_{eff} & = \upsilon_{\n-1} \e^{\n\beta_{\cc}} \frac{m_{\D}^{2}}{m_{\Pl}^{2}} \int\limits_{u_{\text{min}}}^{u_{\text{max}}} \Bigl(f\bigl(R_{\n}(u)\bigr) - \bigl(\varphi^{\prime}(u)\bigr)^2 m^{2}_{\D} \e^{-2\beta_{\cc}} - 2V\bigl( \varphi(u) \bigr)\Bigr) \, r^{\n-1} (u)\, du .
\end{align}
Let us simplify expression \eqref{Lambda_eff_deformed}.
To this end, we integrate equation for $(tt)=(x^i x^i)${--}components over the compact extra space, i.e., $\int d^\n x \sqrt{|g_\n|}$ \eqref{eq_tt_deformed_r}:
\begin{equation}\label{intL}
\begin{gathered}
\int
\Biggl( \biggl(\bigl(R^{\prime}\bigr)^2 f_{RRR} +\Bigl(R^{\prime \prime} + \bigl(\n-1 \bigr) \dfrac{r^\prime}{r}\,R^{\prime} \Bigr) f_{RR} \biggr)m^{2}_{\D}\e^{-2\beta_{\cc}} + \\
 \qquad \qquad \qquad \left. + \, 3H^2f_R - \dfrac{f(R)}{2} + \dfrac{\bigl(\varphi^{\prime}\bigr)^2}{2} m^{2}_{\D}\e^{-2\beta_{\cc}} + V\bigl( \varphi \bigr) \right)  \, r^{\n-1} (u) \, du =0 \end{gathered} 
 \end{equation}
and notice that the first line of \eqref{intL} is the total derivative under the integral and can be omitted by integrating over the closed manifold $M_\n$. Comparison \eqref{Lambda_eff_deformed} and \eqref{intL} gives
\begin{equation}\label{ceff}
c_{eff} = \upsilon_{\n-1} \e^{\n\beta_{\cc}} \frac{m_{\D}^{2}}{m_{\Pl}^{2}}\int\limits_{u_{\text{min}}}^{u_{\text{max}}} \Bigl(6H^2 f_R\bigl(R_\D(u) \bigr) - f\bigl(R_\D(u) \bigr) + f\bigl(R_\n(u) \bigr) \Bigr)\, r^{\n-1} (u) \, du,
\end{equation}
where $f_R\bigl(R_\D(u) \bigr)=2aR_\D(u) + 1$ and $R_\D(u) = 12H^2 + R_\n(u)$ so that $c_{eff}\to 0$ if $H\to 0$, that looks reasonable for the effective action \eqref{S_eff2}.
By expanding the integrand in \eqref{ceff}, one obtains expression
\begin{equation} \label{ceff_quadratic}
c_{eff}= -6H^2 \, \upsilon_{\n-1} \e^{\n\beta_{\cc}} \frac{m_{\D}^{2}}{m_{\Pl}^{2}}\int\limits_{u_{\text{min}}}^{u_{\text{max}}}f_R\bigl(R_{\n}(u)\bigr)\, r^{\n-1} (u) \, du, \end{equation}
which, being combined with \eqref{m_pl_eff_deformed}, yields the well-known relation $H^2=\Lambda/3$ for the standard notations $c_{eff}=-2\Lambda$, where $\Lambda$ is the cosmological constant. For a more general case of the 
$f(R)$ function, terms proportional to $H^6$ and other nontrivial terms can appear in expression \eqref{ceff}. In this case, the  inflationary dynamic requires a separate investigation. Various extensions of the Starobinsky model, or $f(R)$ models in general, are intensively studied in the literature (see the Introduction and references therein), though many of them are significantly limited by experimental data. 
Nonetheless, in our case of quadratic gravity, the terms $H^4$ and higher orders are shown to be absent, and the original Starobinsky model is restored.

As discussed in the previous section, the first stage is finished by the 4-dimensional de-Sitter metric and maximally symmetric extra space. The effective parameter $c_{eff}$ defined by \eqref{coeff_eff_symm} is not equal zero that contradicts observations. 
As follows from more general form \eqref{Lambda_eff_deformed} for $c_{eff}$, it depends on the extra space metric and the scalar field distribution at the inflationary stage. The following subsection demonstrates the existence of an inhomogeneous metric that provides the condition $c_{eff} \simeq 0$ and is responsible for the inflationary stage. Also, the effective parameter $a_{eff} \simeq a_{\text{Starob}} \sim 10^9m^{-2}_{\Pl}$ at the Hubble parameter $H\sim 10^{13}$ GeV. This means that the Starobinsky model is reproduced within a specific pocket universe. According to the discussion above,  it is quantum fluctuations that break the stationary maximally symmetrical extra metric and lead to an inhomogeneous extra metric specific to each pocket universe.


\subsection{Metrics of extra dimensions. Numerical simulations}

Here we discuss possible forms of the extra{-}dimensional metric obtained by numerical solution of a system of equations \eqref{constraint_simple}, \eqref{eq_thetatheta_deformed_r},  \eqref{eq_scalar_deformed_r} and a combination \eqref{eq_tt_deformed_r}$+$\eqref{eq_phiphi_deformed_r}$+$\eqref{Ricci_n_dim_deformed_r}$\cdot f_R / 2\bigl(\n-1 \bigr)$:
\begin{align}\label{comb}
&\Biggl(2\bigl(R^{\prime}\bigr)^2 f_{RRR} \, + \biggr(2R^{\prime \prime}+\bigl(2\n-3 \bigr)\dfrac{r^\prime}{r}\,R^{\prime} \biggl)f_{RR}\Biggr)m^{2}_{\D}\e^{-2\beta_{\cc}} - \\ \nonumber
& \qquad - \dfrac{1}{2}\left(\dfrac{(\n-2)}{r^2}\Bigl(\bigl(r^{\prime}\bigr)^2 -1 \Bigr)  m^{2}_{\D}\e^{-2\beta_{\cc}} - \dfrac{\bigl(6(\n-3)H^2 + R \bigr)}{(\n-1 )} \right)f_R  - \,f(R) + \bigl(\varphi^{\prime}\bigr)^2 m^{2}_{\D} \e^{-2\beta_{\cc}} +2 V\bigl( \varphi \bigr) =0,
\end{align}
resolved with respect to $H$ and unknown functions $r(u)$, $\varphi(u)$ and $R(u)$.
As for boundary conditions, we suppose that $u_{\text{min}} = 0$ is the regular center and
\begin{align}\label{cond}
 r(u_{\text{min}})=0, \quad r^\prime(u_{\text{min}})&=1,\quad  R(u_{\text{min}})=R_0,  \quad \varphi(u_{\text{min}}) = \varphi_0, \\
\label{prime}
R^\prime(u_{\text{min}}) &= 0, \,\,\,\,\, \varphi^\prime(u_{\text{min}})=0.
\end{align}
Conditions \eqref{prime} lead from equations \eqref{eq_tt_deformed_r}, \eqref{eq_scalar_deformed_r} in the limit $u \rightarrow u_{\text{min}} = 0$, provided $f_{RR}(R_0)\neq 0$. The unknown Hubble parameter $H$  can be obtained from \eqref{constraint_simple} applied at the point $u_{\text{min}}=0$ for the particular values of $R_0$ and $\varphi_0$. 

\begin{figure}[!th]
\centering
\begin{subfigure}[t]{0.31\linewidth}
\includegraphics[width=\linewidth]{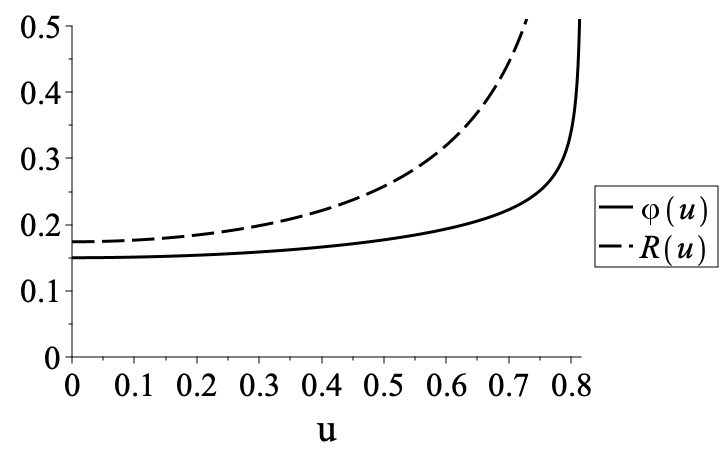}
\includegraphics[width=\linewidth]{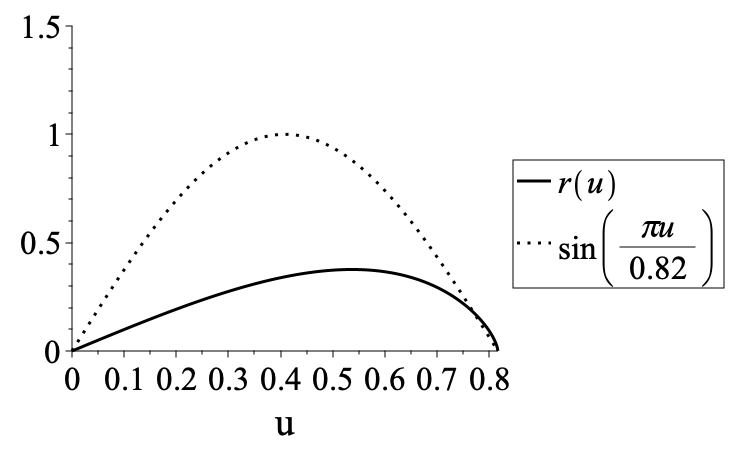}
\caption{: $a=100$, $c=-0.05$, $\n=3$, $m=0.1$, $\beta_{\cc}=2.95$, $H=0.085$, $\varphi_0 = 0.05$ and $R_0 \simeq 0.174$.}
\end{subfigure}
\hspace{0.25cm}
\begin{subfigure}[t]{0.31\linewidth}
\includegraphics[width=\linewidth]{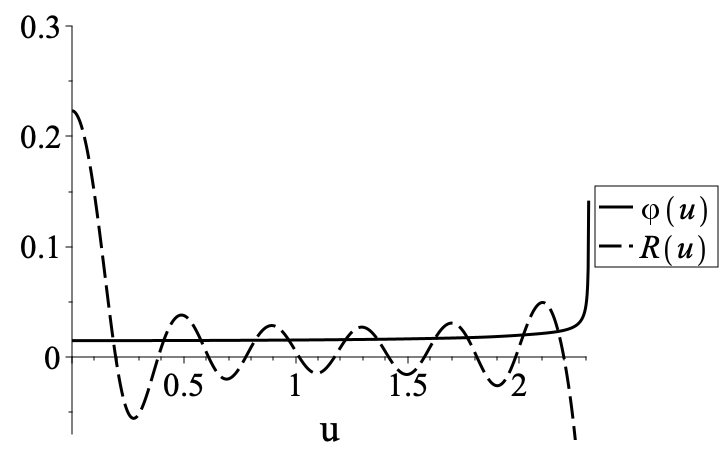}
\includegraphics[width=\linewidth]{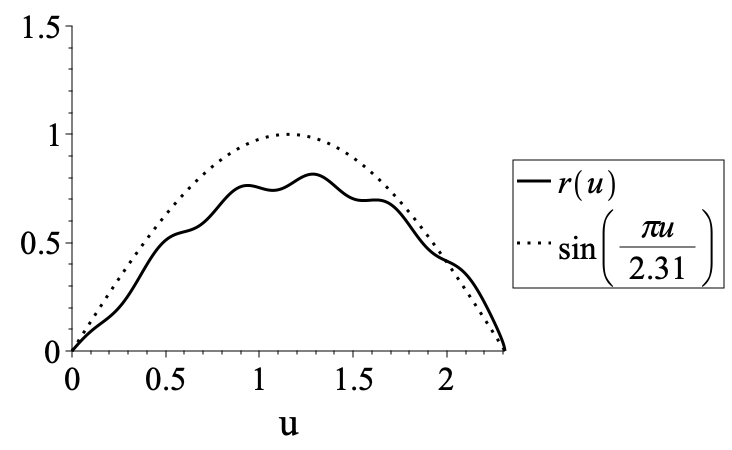}
\caption{: $a=-1.5$, $c=-0.005$, $\n=3$, $m=0.01$, $\beta_{\cc}=3.80$, $H=0.018$, $\varphi_0 = 0.015$ and $R_0 \simeq 0.223$.}
\end{subfigure}
\hspace{0.25cm}
\begin{subfigure}[t]{0.31\linewidth}
\includegraphics[width=\linewidth]{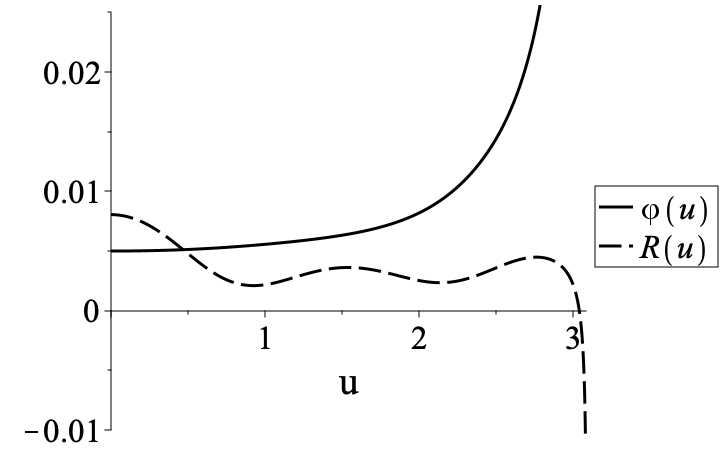}
\includegraphics[width=\linewidth]{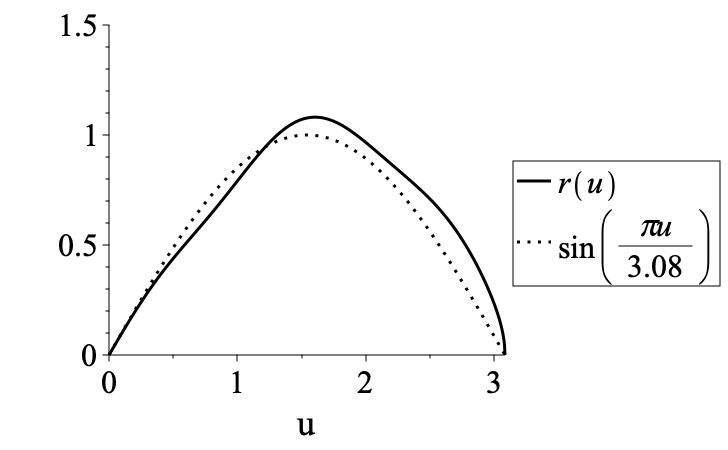}
\caption{:~$a=-45$,~$c=-0.0021$,~$\n=4$, $m=0.01$, $\beta_{\cc}=4.48$, $H=0.011$, $\varphi_0 = 0.005$ and $R_0 \simeq 0.008$.}
\end{subfigure}
\caption{Numerical solution for the functions $r(u)$, $\varphi(u)$, $R(u)$ the given parameters and boundary conditions $r(0)=0$, $r^\prime(0)=1$, $R^\prime(0)=0$, $\varphi(0) = \varphi_0$, $\varphi^\prime(0)=0$ and $R(0)=R_0$ is the~positive~root~of~\eqref{constraint_simple}.} 
\label{Rv_rsin1}
\end{figure}

\begin{figure}[!th]
\centering
\begin{subfigure}[t]{0.31\linewidth}
\includegraphics[width=\linewidth]{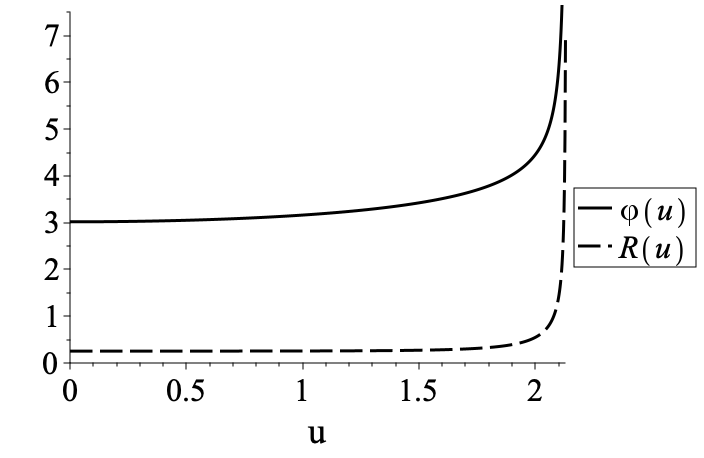}
\includegraphics[width=\linewidth]{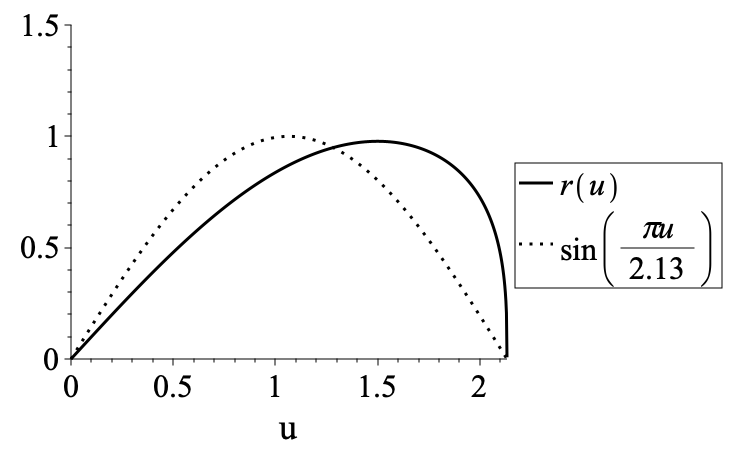}
\caption{: $a=20$, $c=-0.95$, $\n=6$, $m=0.05$, $\beta_{\cc}=2.65$, $H=0.092$, $ \varphi_0 = 3.02$ and $R_0 \simeq 0.254$.}
\end{subfigure}
\hspace{0.5cm}
\begin{subfigure}[t]{0.31\linewidth}
\includegraphics[width=\linewidth]{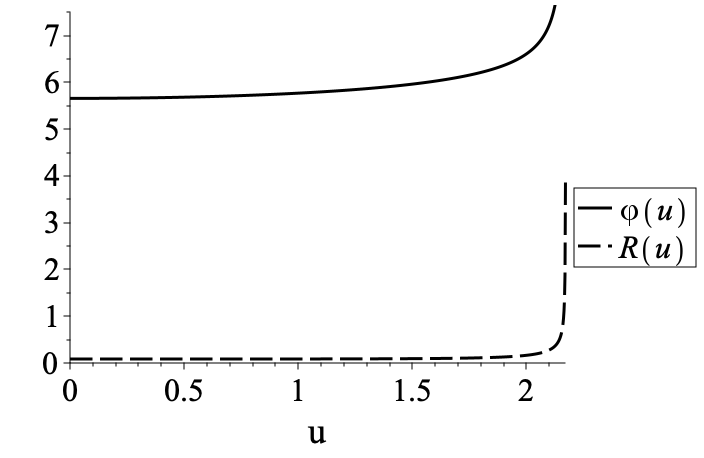}
\includegraphics[width=\linewidth]{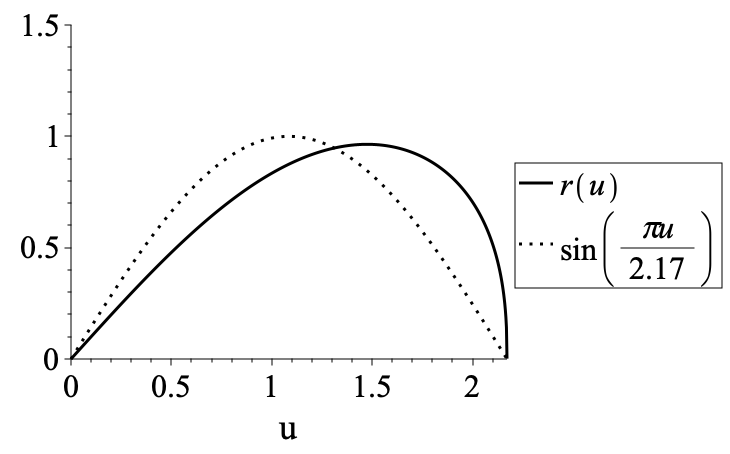}
\caption{: $a=50$, $c=-0.25$, $\n=5$, $m=0.02$, $\beta_{\cc}=3.04$, $H=0.056$, $ \varphi_0 = 5.66$ and $R_0 \simeq 0.085$.}
\end{subfigure}
\caption{Numerical solution for the functions $r(u)$, $\varphi(u)$, $R(u)$ the given parameters and boundary conditions $r(0)=0$, $r^\prime(0)=1$, $R^\prime(0)=0$, $ \varphi(0) = \varphi_0$, $\varphi^\prime(0)=0$ and $R(0)=R_0$ is the~positive~root~of~\eqref{constraint_simple}.} 
\label{Rv_rsin4}
\end{figure}

Some examples of numerical solutions for $\n=3,4,5,6$ and potential $V(\varphi)=\frac{1}{2}m^2 \varphi^2$ are represented in Figures \ref{Rv_rsin1}{--}\ref{Rv_rsin4}, all parameters and boundary conditions are given in units of $m_\D = 1$. The properties of inhomogeneous extra space vary significantly depending on the parameter values. One can see that the Ricci scalar may change its sign, as can be seen from Figures \ref{Rv_rsin1}(b), \ref{Rv_rsin1}(c). We do not dwell on the discussion of the solutions found since, for the purposes pursued in this paper, we use solutions of the form in Figure~\ref{Rv_rsin4}. These solutions have to be averaged over distances $\sim\e^{-\beta_\cc}$, as discussed in the Appendix.

\subsection{Fitting to observational data}
We assume that it is the second step of the space expansion, inflation, that is responsible for such observable parameters as the power spectra of scalar curvature perturbations and tensor perturbations. The parametrization of their power spectra is effectively determined by three parameters: the scalar power spectrum amplitude $A_s$, the scalar spectral index $n_s$ and the tensor-to-scalar power ratio $r$. The Planck 2018 TT, TE, EE+ lowE+lensing data with a combination of BICEP/Keck Array 2018 \cite{2021PhRvL.127o1301A}, baryon acoustic oscillation \cite{2021PhRvD.103h3533A}, constraint the values \cite{2020A&A...641A..10P,2022PhRvD.105h3524T}
\begin{equation}\label{obs}
    n_s = 0.9649 \pm  0.0042 \, , \quad r < 0.032.
\end{equation}
Our model should not contradict these observational limits. The most economic way to find appropriate parameters is to notice that the structure of our effective action \eqref{S_eff2} coincides with that of the Starobinsky model. The latter is known to fit observations quite well. Therefore, it remains only to match the parameters of the Starobinsky model with the parameters of our model. The inflationary predictions originally calculated for Starobinsky's model to the lowest order \cite{1981ZhPmR..33..549M}
\begin{equation}\label{observ}
    n_s - 1 \simeq -\frac{2}{N_\e} \quad \text{and} \quad r \simeq \dfrac{12}{N_\e^2}
\end{equation}
are in good agreement with experimental data for the number of e-folds in the range of $50 < N_\e < 60$. The $R^2$ multiplier has been obtained from the COBE normalization \cite{2020A&A...641A..10P}, as 
\begin{equation}
a_{\text{Starob}} \simeq 1.12 \cdot 10^9 \left(\dfrac{N_\e}{60}\right)^2m^{-2}_{\Pl}. \end{equation}

Our analysis shows that the parameter values of our model $a=20 \, m^{-2}_\D$, $c=-0.95 \, m^{2}_\D$, $\n=6$, $m=0.05 \, m_\D$ suit our aims, and the value of $m_\D$ is discussed below. The numerical selection of an appropriate metric is as follows. Firstly, we should choose additional (boundary) conditions \eqref{cond}, \eqref{prime}. The latter are related by constraint equation \eqref{constraint_simple} and we need to fix only one value $\varphi_0$ or $R_0$. Our choice is $\varphi_0$, the variation of which gives a continuum set of solutions to system \eqref{Ricci_n_dim_deformed_r}-\eqref{eq_scalar_deformed_r}. The second step is to find the exact extra metric which leads to the observable parameters \eqref{obs} or to the parameters used in the Starobinsky model, i.e., the effective value of $a_{eff} \simeq a_{\text{Starob}} \sim 10^9 m^{-2}_{\Pl}$ and a negligibly small $c_{eff}$.

\begin{figure}[!th]
\centering
\begin{subfigure}[t]{0.45\linewidth}
\includegraphics[width=\linewidth]{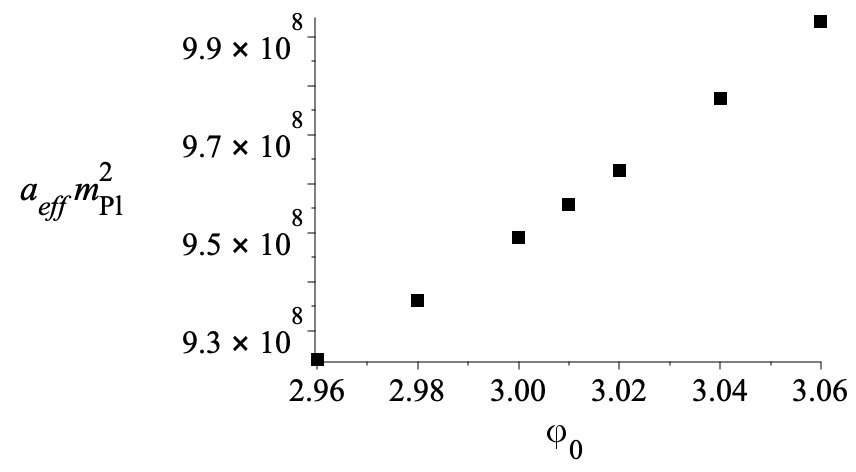}
\includegraphics[width=\linewidth]{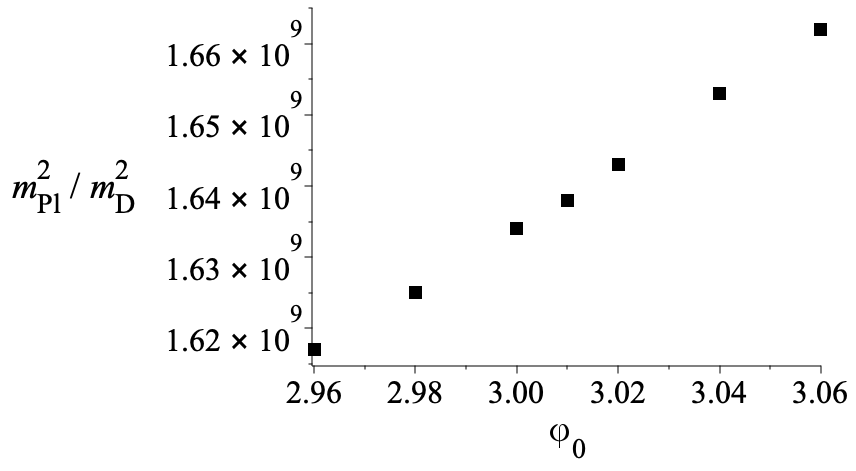}
\includegraphics[width=\linewidth]{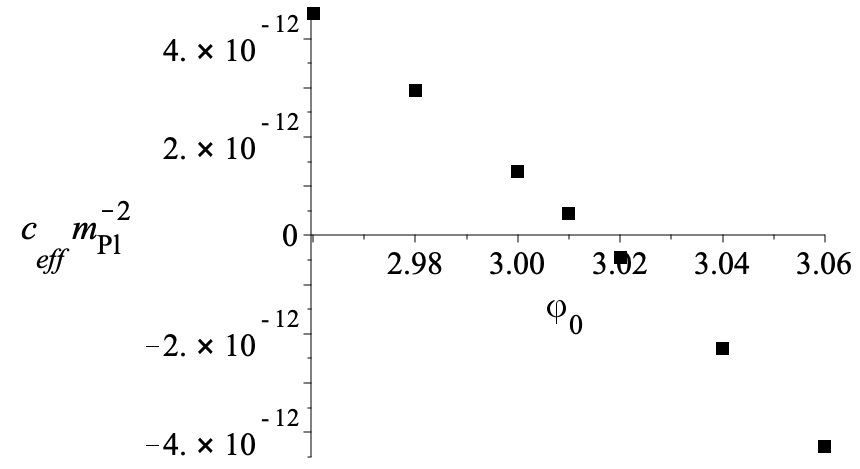}
\vskip1.5mm 
\caption{: $a=20m^{-2}_\D$, $c=-0.95m^{2}_\D$, $\n=6$, \\ $m=0.05m_\D$, $\beta_{\cc}=2.65$, $H=0.092m_\D$.}
\end{subfigure}
\hspace{0.65cm}
\begin{subfigure}[t]{0.45\linewidth}
\includegraphics[width=\linewidth]{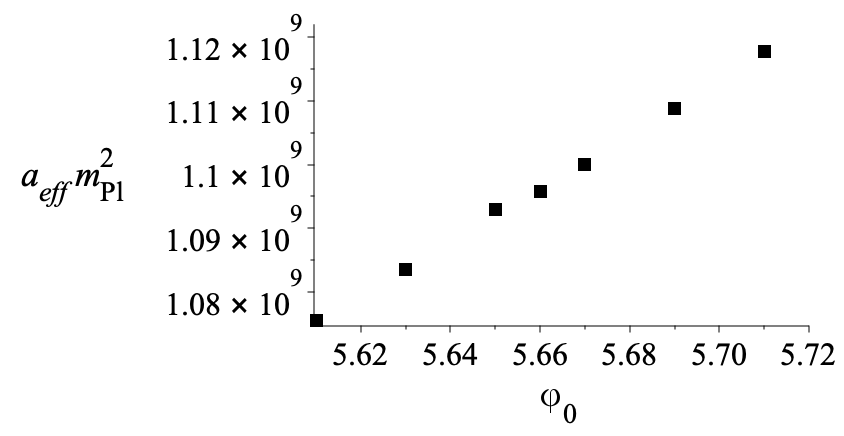}
\includegraphics[width=\linewidth]{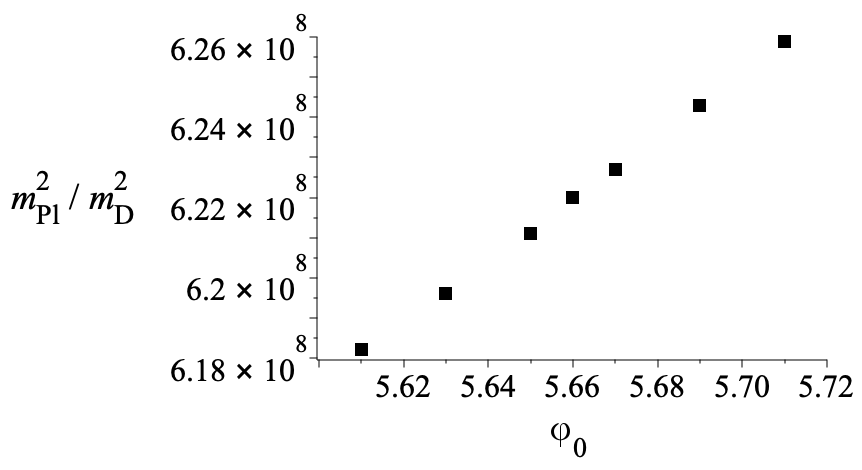}
\includegraphics[width=\linewidth]{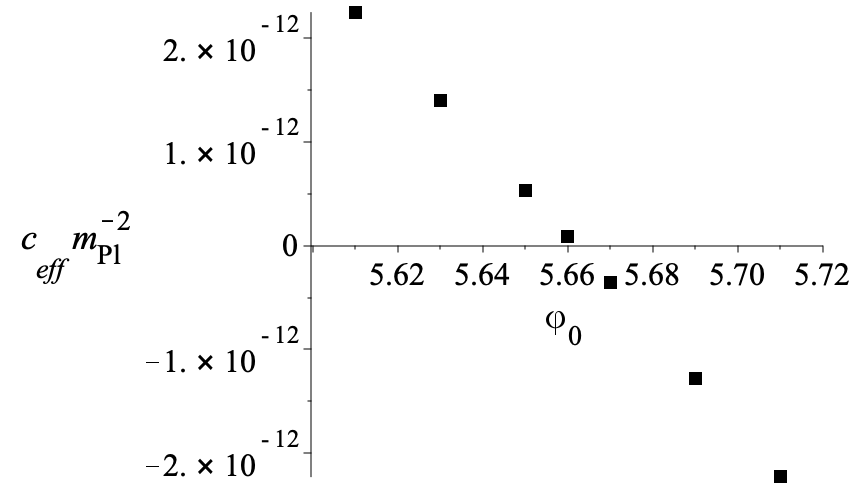}
\caption{: $a=50m^{-2}_\D$, $c=-0.25m^{2}_\D$, $\n=5$, \\ $m=0.02m_\D$, $\beta_{\cc}=3.04$, $H=0.056 m_\D$.}
\end{subfigure}
\caption{The dependence of the effective coefficients \eqref{a_eff_deformed}, \eqref{Lambda_eff_deformed} and the ratio of the 4-dimensional Planck mass to the $\D$-dimensional Planck mass \eqref{m_pl_eff_deformed}  on the boundary value of the scalar field $\varphi_0$ for the given parameters with boundary conditions $r(0)=0$, $r^\prime(0)=1$, $ \varphi(0) = \varphi_0$, $\varphi^\prime(0)=0$,  $R^\prime(0)=0$ and $R(0)=R_0$ is root of~\eqref{constraint_simple}. An example of a solution for such a set of parameters is shown in Figure \ref{Rv_rsin4} above.} 
\label{a_eff_m_pl_c_eff}
\end{figure}

Figure \ref{a_eff_m_pl_c_eff}(a) depicts the variation of the effective parameters as a function of $\varphi_0$.
The shape of the metric for such parameters is presented in Figure \ref{Rv_rsin4}(a). Keep in mind that the parameter $\beta_\cc$ is responsible only for redefining the $u$ coordinate, according to metric \eqref{metric_deformed_r}. The parameter $c_{eff}$ changes its sign somewhere near $\varphi_0 \simeq 3.015$. This means that we can find an extra metric for which $c_{eff} \simeq 0$ with arbitrary good accuracy. It allows one to calculate numerically other effective parameters, i.e., $a_{eff}$ using \eqref{a_eff_deformed} and the ratio of the 4-dimensional Planck mass to the D-dimensional Planck mass from \eqref{m_pl_eff_deformed}. As a result, we have the $\D$-dimensional mass\footnote{For $m_{\Pl}=M_{\Pl}/\sqrt{8\pi} = 2.4 \cdot 10^{18}$ GeV.} $m_\D \sim 10^{14}$ GeV, the Hubble parameter $H \sim 10^{13}$~GeV and $a_{eff} \sim 10^9  m^{-2}_{\Pl}$. This means that the Starobinsky model is restored, and the values of the initial parameters have a reasonable deviation from unity. 

Note that the freedom of choice in the parameter values still remains. We can also reach suitable effective values for other sets of parameters, including another dimension of the extra space. For example, the set $a=50\, m^{-2}_\D$, $c=-0.25 \, m^{2}_\D$, $\n=5$, $m=0.02\, m_\D$ also reproduces the Starobinsky model \eqref{S_eff2} with appropriate parameters $a_{eff} \sim 10^9  m^{-2}_{\Pl}$ and $ c_{eff} \simeq 0$, as shown in Figure~\ref{a_eff_m_pl_c_eff}(b) for the metric in Figure~\ref{Rv_rsin4}(b). Such parameter values give rise to the slow rolling inflation at the Hubble parameter $H\sim 10^{13}$ GeV. It validates our assumption made at the beginning of Section \ref{LES}. 

\begin{figure}[!th]
\centering
\begin{subfigure}[t]{0.288\linewidth}
\includegraphics[width=\linewidth]{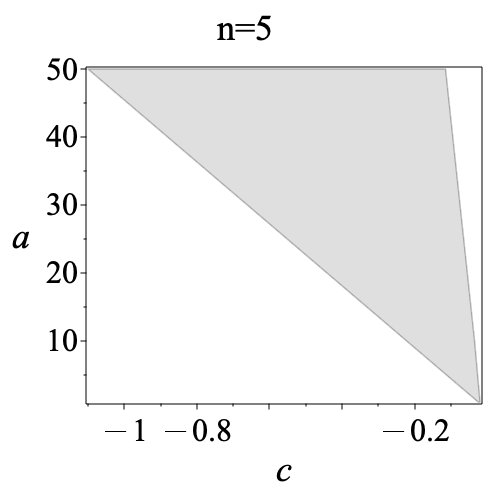}
\vskip1.5mm 
\end{subfigure}
\hspace{0.45cm}
\begin{subfigure}[t]{0.291\linewidth}
\includegraphics[width=\linewidth]{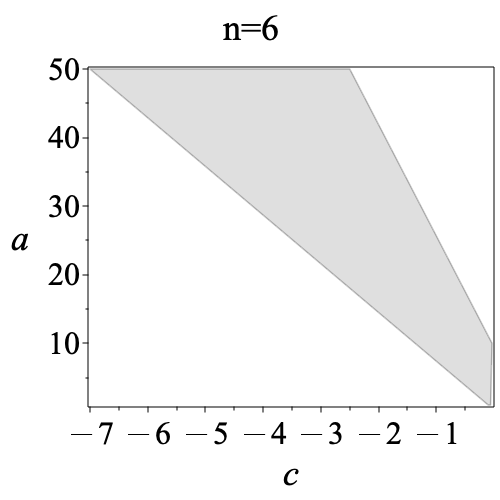}
\vskip1.5mm 
\end{subfigure}
\hspace{0.45cm}
\begin{subfigure}[t]{0.291\linewidth}
\includegraphics[width=\linewidth]{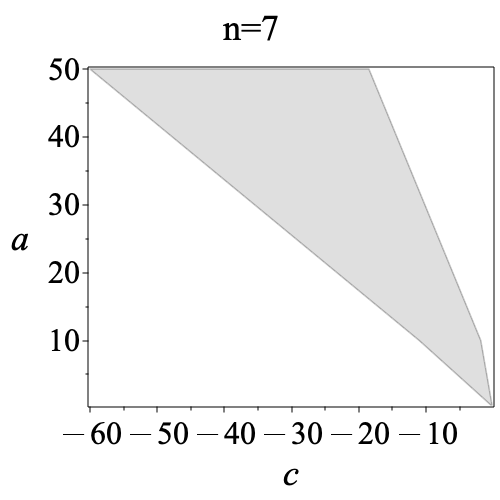}
\end{subfigure}
\caption{Grey regions represent initial Lagrangian parameters $a$ and $c$ in $m_\D=1$ units that ensure $a_{eff} \sim 10^9 m^{-2}_{\Pl} $, $m_\D \sim 10^{14}$ GeV, and $H \sim 10^{13}$ GeV for a given dimension of subspace $\n$.}
\label{Parameters_allowed}
\end{figure}

Figure \ref{Parameters_allowed} shows some acceptable ranges for various number of the extra dimensions.  As one can see, the number of extra dimensions influences mostly the range of the parameter $c$. The choice of the remaining parameters, such as $m$ and boundary conditions\footnote{Note that for real $R_0$ solutions to \eqref{integrable} at the boundary, $m^2 \varphi_0^2 < 4\bigl(\n+3\bigr)\Bigl(3a\bigl(\n+3\bigr)H^2+1\Bigr)H^2+c$ must be held.}, $\varphi_0$ or $R_0$, is owing to achieving $c_{eff} \simeq 0$.

\section{Conclusion}\label{Conclusion}
Many inflationary models explain observational data at the cost of using a small parameter to account for the smallness of the Hubble parameter $H\sim 10^{-6}$ expressed in the Planck units. Also, it is implicitly assumed that one of the model parameters related to the cosmological constant is extremely small. In this paper, we elaborate the inflationary model without small parameters. We also show the way to a significant decrease in the cosmological constant.

There are two stages of space expansion. The first one begins at the highest energy scale of the order of the D-dimensional Planck mass $m_\D$, where D-dimensional space-time of the size $\sim m^{-1}_\D$ is formed. All model parameters differ reasonably from unity when working with $m_\D=1$. The first stage is an unobservable one and is finished by a space with the de Sitter 4-dimensional metric at the scale $\sim 10^{14}$ GeV and the zero value of the scalar field. It is characterized by the Hubble parameter and the metric of compact maximally symmetric extra dimensions, which are strictly constants. As usual, the space expands exponentially, producing more and more causally disconnected volume {--} future pocket universes. 

The new element is the fluctuations of the extra space metric. The extra space metric fluctuates along with the scalar field that leads to the deformation of the extra space. As was shown in \cite{2021arXiv210908373R}, such fluctuations could relax into nontrivial static states. Therefore, each pocket universe is characterized by a specific extra space metric and a scalar field distribution. One of these universes turns out to be randomly endowed with the corresponding metric, which allows it to evolve into the observable universe. The second stage starts when an appropriate extra space metric and a scalar field configuration are formed within a pocket universe. The new metric and the classical field distribution change the effective parameters according to \eqref{a_eff_deformed}-\eqref{Lambda_eff_deformed},
and the stage of observable inflation begins. It does not contradict observations if the effective parameters coincide with those taken from the Starobinsky model, i.e., the standard inflationary scenario is restored. 

We show that the extra space metric leading to the zero cosmological constant can be found. In fact, it is enough to make this constant small during inflation because many other unaccounted effects contribute to its value at low energies. In particular, the quantum effects inevitably lead to the energy dependence of Lagrangian parameters. Here we assume that the physical laws are formed at the highest energy scale and that initial parameter values are of the order of this energy scale. The relation between low energy parameter values and high energy ones are discussed in \cite{2018JCAP...12..004G,2018PhRvD..98d3505L}. The quantum corrections caused by matter fields to these parameters defined at the Planck scale are small if coupling constants are small \cite{Rubin:2020pqu}. This means that such quantum effects cannot be responsible for reducing the parameter values by many orders of magnitude - from the Planck scale to the electroweak scale. The classical mechanism discussed in this paper was elaborated just for this aim.

At the same time, gravity is also responsible for quantum corrections to the original Lagrangian. And these corrections are not small at high energies where the parameter values are chosen \cite{2018PhRvD..98d3505L,2022CQGra..39e5003A}. Their calculation based, for example, on the renormalization group flow deserves deep discussion in the future.
Additional limits to the Lagrangian parameters are posed in paper \cite{2019IJMPD..2841004N}. They are the reasons for the fact that the extra space size should be larger than the Planck unit, $r \gg m^{-1}_\D$ (see the discussion in the Appendix). Another inequality $r \ll H^{-1}_{\text{Infl}}$ is necessary if an inflationary model assumes slow rolling. Note that these inequalities give limits $m_\D > H_{\text{Infl}}\sim 10^{13}$~GeV and $r<10^{-27}$~cm that are much stronger than those obtained in the collider experiments. 

Our model of inflation does not contain unacceptably small or large parameters of the Lagrangian. The effective parameters suitable for explanation of the observational data are formed by the inhomogeneous extradimensional metric. The latter, in its turn, is the result of quantum fluctuations.

\section*{Acknowledgements}
The authors are grateful to the anonymous referees for proposing improvements to the text. The work of P.P. was supported by the Foundation for the Advancement of Theoretical Physics and Mathematics “BASIS”, Grant No. 22-1-5-114-1 and the MEPhI Program Priority 2030.
The work of S.G.R. has been supported by the Kazan Federal University Strategic Academic Leadership Program. P.P. is also grateful to M.Valialshchikov for his interest in the work.

\section*{Appendix: Limits caused by quantum fluctuations}
Here we show that restrictions posed by quantum fluctuations have to be taken into account when studying compact extra dimensions for two reasons. The first one is evident {---} the $\D$-dimensional Planck mass $m_\D$ differs from the 4-dimensional one, so the $\D$-dimensional Planck lengths $l_\D=m^{-1}_\D$ could be much larger than the standard one. The classical form of the metric of compact dimensions is irrelevant if their size is smaller than $l_\D$. In our case, it means that the classical solutions discussed above are invalid at the intervals $\delta l= \e^{\beta_\cc}\Delta u < 1$. Therefore, the solution to the classical equations should not be taken into account in the interval $\Delta u\sim  \e^{-\beta_\cc}$. For the same reason, the scale of compact extra dimensions can not be arbitrarily small. Here and below, all values are expressed in terms of the $\D$-dimensional Planck mass $m_\D$. Another point is the critical length $\delta l_\cc$ for which the quantum fluctuations of fields can not be neglected. It is of interest to compare both lengths, $l_\D$ and $\delta l_\cc$.

Suppose that the scalar field moves starting from the configuration $\chi(x,t_1)$ at the moment $t_1$ to the configuration $\chi(x,t_2)$ at the moment $t_2$. Its average value can be found as
\begin{equation}
\chi_\cc(x,t)\equiv Z\int_{\chi(x,t_1)}^{\chi(x,t_2)}{\cal D}[\chi]\cdot \chi(x,t)\cdot \exp \left[{i}S[\chi]\right]
\end{equation}
The function $\chi_\cc(x,t)$ includes all quantum fluctuations and is supposed to be measured by classical instruments. The saddle-point method applied to such integrals gives 
$\chi_\cc =\chi_{\text{cl}}+\delta\chi_{\text{q}}$,
where the classical field $\chi_{\text{cl}}$ is a solution of the equation following from the action minimisation. If the quantum correction $\delta\chi_{\text{q}}$ is neglected, the average value is equal to the classical one. Our aim is to find an interval for which the quantum fluctuations $\delta\chi_{\text{q}}$ are important.

Let us estimate the quantum corrections to the field evolution in a tiny volume $\upsilon_\D\simeq \delta l_\cc^\D$ of $\D$-dimensional space. The path integral acting in the Euclidean space 
\begin{eqnarray}
&&K\equiv \int{\cal D}[\chi]\cdot \exp [{-}S_{\upsilon_\D}[\chi]]=\exp [{-}S[\chi_{\text{cl}}]]\int{\cal D}[\chi]\cdot \exp [{-}S_{\upsilon_\D}[\chi_{\text{cl}}+\delta\chi_{\text{q}}]]
\end{eqnarray}
describes the transition amplitude in the small volume. This smallness is used in the decomposition
\begin{equation}\label{SvD}
S_{\upsilon_\D}[\chi_{\text{cl}}+\delta\chi_{\text{q}}]\simeq S_{\upsilon_\D}[\chi_{\text{cl}}]+ \frac {\upsilon_\D}{2}\left( \frac{\delta \chi_{\text{q}}^2}{\delta l_\cc^2}+m^2 \delta \chi_{\text{q}}^2 \right) + S_{\text{q}}[\delta \chi_{\text{q}}]+...;\quad S_{\text{q}}[\delta \chi_{\text{q}}]=\upsilon_\D\lambda \delta \chi_{\text{q}}^\alpha, \quad \alpha=3,4,5...
\end{equation}
Here we put $\partial^{\M} \chi \partial_{\M} \chi \simeq \cfrac{\delta \chi^2}{\delta l^2}$. The term $S_{\text{q}}$ contains higher degrees of the scalar field which is responsible for the quantum effects. The saddle-point method holds if the quadratic term dominates. In the saddle-point method, the quadratic term in \eqref{SvD} contributes if
\begin{equation}\label{qq}
\frac{\upsilon_\D}{2}\left( \frac{\delta \chi_{\text{q}}^2}{\delta l_\cc^2}+m^2 \delta \chi_{\text{q}}^2 \right) \leq 1
\quad \text{or} \quad 
\delta \chi_{\text{q}} \leq \chi_{\text{bound}}\equiv  \left( 1+m^2 \delta l_\cc^2 \right)^{-1/2}\delta l_\cc^{\D/2-1}.
\end{equation}
The quantum corrections start to dominate if the last term in \eqref{SvD} is larger or equal unity. Together with~\eqref{qq}, it gives approximate equation
\begin{equation}\label{Sq}
S_{\text{q}}[\chi_{\text{bound}}]\sim \lambda \chi_{\text{bound}}^\alpha\sim \lambda\left( 1+m^2 \delta l_\cc^2 \right)^{-\alpha/2}\delta l_\cc^{\alpha(\D/2-1) } \sim 1,\quad 
\end{equation}
which can be used to find the length where the classical solutions for the scalar field are irrelevant.

We accept the equality $l_\D\simeq 1=m^{-1}_\D$ as the limit of the classical treatment in this paper.
\printbibliography[title={References}, heading=bibintoc]
\end{document}